\documentclass[sigconf]{acmart}

\usepackage{tabularx}
\usepackage{subcaption}
\usepackage{algorithm}
\usepackage{algpseudocode}
\algtext*{EndIf} 
\algtext*{EndFor} 
\algtext*{EndWhile}

\usepackage{pifont}
\usepackage{float}
\usepackage{graphicx}
\usepackage{enumitem}

\DeclareMathOperator*{\argmin}{arg\,min}

\AtBeginDocument{%
  \providecommand\BibTeX{{%
    \normalfont B\kern-0.5em{\scshape i\kern-0.25em b}\kern-0.8em\TeX}}}

\setcopyright{acmlicensed}
\copyrightyear{2018}
\acmYear{2018}
\acmDOI{XXXXXXX.XXXXXXX}

\acmConference[SIGMOD '25]{In Proceedings of the 2025 International Conference on Management of Data}{June 22--27,2025}{Berlin, Germany}
\acmISBN{978-1-4503-XXXX-X/18/06}

\newcommand*\justify{%
  \fontdimen2\font=0.4em
  \fontdimen3\font=0.2em
  \fontdimen4\font=0.1em
  \fontdimen7\font=0.1em
  \hyphenchar\font=`\-
}

\renewcommand{\texttt}[1]{%
  \begingroup
  \ttfamily
  \begingroup\lccode`~=`/\lowercase{\endgroup\def~}{/\discretionary{}{}{}}%
  \begingroup\lccode`~=`[\lowercase{\endgroup\def~}{[\discretionary{}{}{}}%
  \begingroup\lccode`~=`.\lowercase{\endgroup\def~}{.\discretionary{}{}{}}%
  \catcode`/=\active\catcode`[=\active\catcode`.=\active
  \justify\scantokens{#1\noexpand}%
  \endgroup
}

\begin{document}
\sloppy


\title{iRangeGraph: Improvising Range-dedicated Graphs for Range-filtering Nearest Neighbor Search}



\author{Yuexuan Xu\textsuperscript{*}}
\affiliation{%
  \institution{Nanyang Technological University}
  \country{Singapore}}
\email{yuexuan001@e.ntu.edu.sg}

\author{Jianyang Gao\textsuperscript{*}}
\affiliation{%
  \institution{Nanyang Technological University}
  \country{Singapore}}
\email{jianyang.gao@ntu.edu.sg}

\author{Yutong Gou}
\affiliation{%
  \institution{Nanyang Technological University}
  \country{Singapore}}
\email{yutong003@e.ntu.edu.sg}

\author{Cheng Long\textsuperscript{\textdagger}}
\affiliation{%
  \institution{Nanyang Technological University}
  \country{Singapore}}
\email{c.long@ntu.edu.sg}

\author{Christian S. Jensen}
\affiliation{%
  \institution{Aalborg University}
  \city{Aalborg}
  \country{Denmark}}
\email{csj@cs.aau.dk}

\thanks{\textsuperscript{*}Both authors contributed equally to this research.}
\thanks{\textsuperscript{\textdagger}Corresponding author.}







\renewcommand{\shortauthors}{Xu and Gao, et al.}

\begin{abstract}

Range-filtering approximate nearest neighbor (RFANN) search is attracting increasing attention in academia and industry. Given a set of data objects, each being a pair of a high-dimensional vector and a numeric value, an RFANN query with a vector and a numeric range as parameters returns the data object whose numeric value is in the query range and whose vector is nearest to the query vector. 
To process this query, a recent study proposes to build $O(n^2)$ dedicated graph-based indexes for all possible query ranges to enable efficient processing on a database of $n$ objects. As storing all these indexes 
is prohibitively expensive, the study constructs compressed indexes instead, which 
reduces the memory consumption considerably.
However, this incurs suboptimal performance because the compression is lossy. In this study, instead of materializing a compressed index for every possible query range in preparation for querying, we materialize graph-based indexes, called elemental graphs, for a moderate number of ranges. We then provide an effective and efficient algorithm that during querying can construct an index for any query range using the elemental graphs. We prove that the time needed to construct such an index is low.  We also cover an experimental study on real-world datasets that provides evidence that the materialized elemental graphs only consume moderate space and that the proposed method is capable of superior and stable query performance across different query workloads.

\end{abstract}


\maketitle

\section{introduction}
\label{sec:intro}
Nearest neighbor (NN) search in high-dimensional Euclidean spaces occurs in a broad range of settings, including in information retrieval~\cite{liu2007survey}, and machine learning systems (e.g., retrieval-augmented generative AI)~\cite{lewis2020retrieval} and in database systems~\cite{wang2021milvus}. 
The curse of dimensionality reduces the utility of exact NN  search on large datasets, primarily due to long response time. To  obtain better response time at the cost of reduced accuracy, approximate nearest neighbor (ANN) search~\cite{graph_benchmark, hnsw,nsg,annbenchmark, li2019approximate, indyk1998approximate} had received attention. 
Among existing ANN methods, graph-based methods have shown promising performance in terms of the trade-off between response time and result accuracy in in-memory settings~\cite{fanng, nsw, fu2016efanna, iwasaki2016pruned, spann, rng, graph_benchmark, hnsw,nsg,2019diskann,li2019approximate,ChenW18,nssg}.

There has been a recent growing interest in academia~\cite{patel2024acorn,WindowFilter} and industry (including Apple~\cite{HQI}, Zilliz~\cite{wang2021milvus}, and Alibaba~\cite{wei2020analyticdb}) in supporting ANN queries that also involve constraints on numerical data attributes.
Here, a data object consists of two components: (1) a vector and (2) one or more numeric attribute values. A query then takes two parameters: a query vector and constraints on the  numeric attribute values. 
In this paper, we study the query where a data object has one numeric attribute value and the constraint is a range, called the range-filtering ANN (RFANN) query~\cite{segmentgraph,WindowFilter}. 
As mentioned in industrial papers~\cite{HQI,wang2021milvus,wei2020analyticdb}, this query is important in many real-world systems. For example, on an e-commerce platform, a user could upload an image of a product, encoded as a vector, to find similar products within a certain price range. A data object is \textit{in-range} if its numeric attribute value is in the query range; otherwise, it is \textit{out-of-range}.

Three strategies that differ in when the range filtering is done can be adopted to process RFANN queries. \texttt{Pre-filtering} first eliminates out-of-range data objects and finds the nearest neighbor of the query vector among the remaining objects that are not indexed.
\texttt{Post-filtering} first conducts (graph-based) ANN search to find data objects whose vectors are close to the query vector and then eliminates any out-of-range data objects.
\texttt{In-filtering} integrates the range filtering into the (graph-based) ANN search process that then only visits in-range objects (for details, see Section~\ref{subsec:RFANNS}).
While these strategies are intuitive, they have the inherent issue that they are each efficient for certain query workloads only. For example, when the predicate is unselective (i.e., a large fraction of data objects have the attribute value within the query range), \texttt{Pre-filtering} degenerates to sequential search on a large dataset.
\texttt{Post-filtering} and \texttt{In-filtering} are suboptimal when the predicate is very selective.
In particular, \texttt{Post-filtering} will visit many out-of-range objects in order to find the nearest neighbor.
As for \texttt{In-filtering}, as it only visits in-range objects during the graph-based ANN search, the nearest neighbor might be unreachable during the search process, which reduces search performance.

Most existing proposals~\cite{HQI,wei2020analyticdb,wang2021milvus,segmentgraph,WindowFilter} address the shortcomings of the three strategies.
Milvus~\cite{wang2021milvus} and ADBV~\cite{wei2020analyticdb} propose to automatically select a strategy on a per query basis using a cost model. 
Although this avoids using the strategies where they are distinctively disadvantageous, the inherent issues of the strategies have yet to be fully resolved. 
In particular, especially when the predicate is neither extremely selective or unselective, no matter which strategy is adopted, the inherent issues remain, causing suboptimal performance.

To address the inherent issues of the basic strategies, a recent study~\cite{segmentgraph} considers building a dedicated graph-based index for every possible query range. 
To process a query, the index matching the query range is used, thereby reducing the query to an ANN query.
However, as also observed in the study, building an index for each possible query range is not feasible in practice because for a database of $n$ objects, $O(n^2)$ dedicated indexes would be built (each corresponding to an index on a sub-interval of the list of objects sorted wrt their numeric values), resulting in a space complexity of $O(n^3m)$ for storing the indexes. The study thus proposes to compress the $O(n^2)$ indexes, to reduce substantially the memory consumption, yielding an empirical memory consumption after compression that is much smaller than $O(n^3m)$.
However, the lossy nature of the compression has the effect of reducing index quality, in turn reducing query performance. 
Thus, as reported in the study~\cite{segmentgraph} and verified in Section~\ref{subsec:exp single attribute}, for queries with short ranges, the method cannot achieve >0.8 recall.

In this paper, we follow the idea of performing search on a dedicated graph-based index for a given query range to avoid the issues inherent to the three basic strategies. 
However, unlike the existing proposal~\cite{segmentgraph} that builds compressed graphs for all possible query ranges, we materialize graphs for only a moderate number of ranges in preparation for querying.
These graphs, called \emph{elemental graphs}, are then used to construct dedicated graphs for any query range seen during querying.
The construction occurs on the fly in the sense that edges of an object are only constructed when having to visit its  neighbors during query processing.
The algorithm for constructing the dedicated graph in the query phase incurs only a low overhead compared to the cost of searching the dedicated graph.
We call this new method, which aims to \underline{i}mprovise \underline{range}-dedicated \underline{graph} indexes during querying, \texttt{iRangeGraph}.
We further extend \texttt{iRangeGraph} to support multi-attribute range-filtering ANN querying.
Moreover, we propose a simple yet effective strategy to further enhance the search performance on the multi-attribute query.

 The paper makes the following main contributions.
\begin{enumerate}
\item We propose the \texttt{iRangeGraph} method for computing the RFANN query. It materializes a moderate number of elemental graphs prior to querying and constructs a dedicated graph index for any given query range based on these elemental graphs on the fly during querying. 
 The overhead of constructing dedicated graph during querying is low, meaning that the RFANN query can be reduced to an ANN query on a dedicated graph index with little overhead. (Section~\ref{sec:methodology})
\item We extend \texttt{iRangeGraph} to support multi-attribute range-filtering ANN queries that involve
conjunctive predicates on multiple numeric attributes. We also propose a simple yet effective strategy to further enhance the performance of the multi-attribute RFANN query. (Section~\ref{sec:MultiAttribute})
\item We report on extensive experiments on real-world datasets.
The observations are summarized as follows. 
(a) In general, our method achieves superior search performance with moderate memory footprint on all the datasets and query workloads for RFANN query, e.g., it outperforms the most competitive baseline method by 2x-5x in qps (query per second) at 0.9 recall with consistently smaller memory footprint. 
(b) The extension to support multi-attribute RFANN queries also exhibits state-of-the-art performance. It outperforms the most competitive baseline by 2x--4x in qps at 0.9 recall. 
(c) We measure the gap in search performance between our method and the dedicated graph-based indexes which are materialized for the given query ranges. While the latter is expected to enable ideal search performance with impractical space consumption, we find that our method is only slower than the ideal performance by less than 2x at 0.9 recall. Thus, the proposed method is capable of performance very close to the ideal performance with much lower space consumption, i.e., $O(nm\log n)$ vs. $O(n^3m)$. (Section~\ref{sec:experiment})
\end{enumerate}

In addition, Section~\ref{sec: motivation} presents the range-filtering ANN query and existing studies, while Section~\ref{sec: related work} covers related work and Section~\ref{sec: conclusion} concludes the paper.

\section{ANN and Range-filtering ANN}
\label{sec: motivation}

\subsection{ANN and Graph-based Algorithms}
\label{subsec:anns_and_graph}

\smallskip\noindent\textbf{Approximate Nearest Neighbor Search.} 
Given a query vector $q$ and a set of data objects where each object has a vector $v$, the nearest neighbor (NN) query finds the object whose vector has the smallest distance to $q$. 
Due to the curse of dimensionality~\cite{indyk1998approximate}, the exact NN query often takes unacceptably long response time to compute.
Thus, it is often relaxed to an approximate NN (ANN) query in order to trade improved response time for reduced result accuracy~\cite{graph_benchmark,wang2022survey,gao2023high,indyk1998approximate, nsg, 2019diskann,hnsw}.
Furthermore, the ANN query is usually extended to return $k$ approximate nearest neighbors.
For ease of presentation, we assume $k=1$ in algorithm descriptions, while we note that all the techniques in this paper are adapted and evaluated for a general $k$ in Section~\ref{sec:experiment}. 
Further, we focus on the in-memory setting, where the vectors and indexes are in RAM~\cite{hnsw, nsg,graph_benchmark, annbenchmark}.

\smallskip\noindent\textbf{Graph-based ANN Methods.} Among 
the proposed ANN algorithms~\cite{2019diskann,gao2023high,hnsw,nsg,addq,pq,pqfs,beygelzimer2006cover,huang2015query}, graph-based methods exhibit superior performance in terms of the time-accuracy trade-off~\cite{wang2022survey,graph_benchmark, annbenchmark, li2019approximate, hnsw,nsg}. 
Prior to querying, these methods all build a graph index on the data objects, but they differ in how they construct the edges in the graph. 
During querying, they perform greedy search on the graph in iterations. Specifically, the algorithms start with an entry point in the graph. Then in each iteration, they identify the data object nearest to the query that has been visited so far, and then visit all its neighbors. 
If the vectors of all the neighbors are further away from the query vector than the currently nearest vector, the algorithm terminates. 
In order to achieve practical accuracy, the algorithm is usually extended to the greedy beam search which tunes a parameter named the beam size $b$ to control the time-accuracy trade-off. 
Specifically, different from the greedy search, the greedy beam search terminates when the neighbors of the first $b$ nearest objects that have been visited so far all have their distances to the query larger than the distances of the first $b$ nearest objects. 
In particular, the greedy search is a special case of the greedy beam search where $b=1$. 
Clearly, a larger beam size will increase the number of the visited data objects before termination, and thus, would lead to better accuracy with higher time cost.

Most of the state-of-the-art graph-based methods, including HNSW~\cite{hnsw}, NSG~\cite{nsg}, and  DiskANN~\cite{2019diskann}~\footnote{ The classical library hnswlib has updated its pruning rules with that of RNG due to its better empirical performance. The pruning rule of DiskANN is generalized from RNG's pruning rule by introducing a new parameter $\alpha$. 
When $\alpha=1$, it is identical to the original RNG's pruning rule.}, construct their graphs by computing an approximate variant of a  so-called RNG~\cite{rng, graph_benchmark, hnsw,nsg,2019diskann,li2019approximate,ChenW18,nssg}. This graph is defined as follows.

\begin{definition}[\textit{RNG}~\cite{rng}]
\label{def:rng}
    Given a set of vectors $\mathcal{D}$, let $(u,v)$ be a pair of vectors in $\mathcal{D}$ and let $\delta(u,v)$ be the distance between the vectors. RNG is a graph where the set of vertices represent vectors in set $\mathcal{D}$ and where edges are included as follows: an edge from $u$ to $v$ is included in the RNG if and only if $\forall u'\in \mathcal{D}$, $u'$ cannot prune the edge $(u,v)$. 
    In particular, $u'$ can prune $(u,v)$ if and only if $u'$ is closer to $u$ than $v$ is and also closer to $v$ than $u$ is, i.e., $\delta(u,u') < \delta (u,v)$ and $\delta(v,u') < \delta (u,v)$. 
\end{definition}
The RNG guarantees that the edges of one data object~\footnote{Without further specification, the edges are all directed in this paper. The edges of a data object refer to its outgoing edges.} that are retained after pruning have diversified directions, i.e., an edge is pruned if there exists a shorter edge that has similar direction to it in the high-dimensional space.
Since computing an RNG graph for $n$ data objects takes $O(n^3)$ time~\cite{2019diskann}, which is prohibitively expensive when $n$ is large, existing methods mostly construct approximate RNG-based graphs~\cite{hnsw,nsg,2019diskann,li2019approximate,ChenW18,nssg}. 
Specifically, when including the edges for a data object, they first search for a few candidates in the database (e.g., approximate nearest neighbors).
Then they apply the pruning rule of RNG to the candidates. The candidates that cannot be pruned are retained, and edges are built from the data object to the candidates. 
To prevent a data object from having too many edges, a maximum out-degree $m$ is imposed. 
According to a recent benchmark~\cite{graph_benchmark}, the RNG-based graphs have shown highly promising ANN query performance on many real-world datasets, although we note that the algorithms for candidate generation differ from method to method.

\subsection{Range-Filtering ANN Query}
\label{subsec:RFANNS}

As already explained in the introduction, RFANN queries, which combine ANN querying on vectors with range querying on numeric attributes, have gained attention in both academia and industry.
We focus on the setting where a data object has a vector and one numeric attribute and where a query retrieves the data object with a vector that is nearest to a query vector and a numeric value that is in a query range. This is the range-filtering ANN (RFANN) query~\cite{WindowFilter,segmentgraph}, which is defined as follows.

\begin{definition}[Range-filtering ANN (RFANN) query~\cite{WindowFilter,segmentgraph}]
Formally, a data object is defined as $O=(v, a)$, where $v$ is a vector and $a$ is a numeric attribute value.
Let $\mathcal{D}=\{O_1,O_2,...,O_n\}$ be the set of the data objects.
We define an RFANN query as $Q = (q, a_l, a_r)$, where $q$ is a vector, and $[a_l, a_r]$ denotes a range.
The RFANN query returns the result $\argmin_{O \in \mathcal{D}, Q.a_l \le O.a \le Q.a_r} \delta(Q.q, O.v) $,  where $\delta$ is a distance function.
\end{definition}

A data object is \textit{in-range} if the value of its numeric attribute is in the query range; otherwise, it is \textit{out-of-range}.
Without loss of generality, we assume that the data objects are given in an ascending order with respect to their numeric attribute values, i.e., $\forall i\leq j$, we have $O_i.a \le O_j.a$. 
Otherwise, we initially sort the dataset and build a one-to-one mapping between the ranking and the attribute values of the objects.
When a query arrives, we can easily find the smallest ranking $L$ and the largest ranking $R$ (using binary search) such that the in-range objects are exactly the objects with indices (rankings) between these two values.
Then the RFANN query is reduced to finding $\argmin_{L \le i\le R} \delta(Q.q, O_i.v) $.
Going forward, we use $[L,R]$, to denote a query range and $i$ to denote $O_i$.
We note that the number of distinct attribute values of the dataset is upper-bounded by the size of the dataset. 
To ease the presentation, we first assume that all data objects have distinct attribute values. At the end of Section~\ref{sec:methodology}, after presenting our method, we present a more detailed discussion of the cases where duplicate attribute values exist.
Thus, after the above operation is performed, the attribute values of the data objects are all mapped to an integer that is no larger than $n$. The raw query ranges are all mapped to a range of indices $[L,R]$, which is a sub-interval of $[1,n]$.
Therefore, in the following sections, we use $n$ to denote both the size of the dataset and the cardinality of the attribute.
In addition, we note that because the attribute values are all mapped to their rankings, the distribution of the attribute values does not impact the performance of an algorithm.
Related work on queries with other types of attributes and constraints is covered in Section~\ref{sec: related work}.

We proceed to revisit the three basic strategies for processing RFANN queries.
\underline{\textbf{\texttt{Pre-filtering}}} first filters out the out-of-range objects (e.g., using binary search)  and then iterates through the remaining objects to find the nearest neighbor. For unselective queries, this degrades to a linear scan of most of the data.
\underline{\textbf{\texttt{Post-filtering}}} first conducts ANN search on the dataset (e.g., with graph-based algorithms) and then finds the nearest neighbor among the in-range objects visited.
Given a highly selective query, this strategy visits many out-of-range objects, which is suboptimal. \underline{\textbf{\texttt{In-filtering}}} integrates the range filtering into the (graph-based) ANN search. 
Specifically, unlike vanilla graph-based methods that visit \textit{all} neighbors~\footnote{Without further specification, by ``neighbors,'' we mean neighbors of a node in a graph index, not the nearest neighbors in high-dimensional space.} of an object during querying (Section~\ref{subsec:anns_and_graph}), the \texttt{In-filtering} strategy only visits the \textit{in-range} neighbors. 
This strategy cannot handle different query ranges with the fixed graph built for ANN search on the whole database.
In particular, when query ranges are small, a data object is likely to have only a few, or even no, in-range neighbors.
If we then increase the number of edges in the graph-based indexes to solve the issue of short query ranges, then when considering queries with long ranges, a data object may have too many in-range neighbors.
Overall, with the \texttt{In-filtering} strategy, the quality of a fixed graph-based index is suboptimal for some query ranges.

Most existing proposals aim to mitigate the issues inherent to the three strategies~\cite{HQI,wei2020analyticdb,wang2021milvus,segmentgraph,WindowFilter}. 
In particular, \texttt{ADBV}~\cite{wei2020analyticdb} and \texttt{Milvus}~\cite{wang2021milvus} use a cost model to automatically select the best strategy for a given query. \texttt{Milvus}~\cite{wang2021milvus} further partitions a dataset into several subsets with consecutive attribute values.
During querying, it conducts RFANN search separately on the subsets that intersect with the query range and then merges their results. 
Another study~\cite{WindowFilter} proposes \texttt{SuperPostfiltering} that presets multiple overlapping ranges and builds a graph-based index for each range separately. 
During querying, it first finds the shortest range that covers the query range. Then it conducts RFANN search on the range with the \texttt{Post-filtering} strategy. 
Although these proposals successfully avoid the distinctively disadvantageous cases of the basic strategies, we note that the inherent issues of the strategies have yet to be fully resolved. 
For example, for a query range of size $s$, \texttt{SuperPostfiltering}~\cite{WindowFilter} performs \texttt{Post-filtering} on a subset of the database  where there are at most $4s$ data objects.
However, among them, there may still be $3s$ out-of-range objects, which causes suboptimal performance.

To address the inherent issues of the basic strategies, a recent study~\cite{segmentgraph} considers building a dedicated graph-based index for each possible query range. 
During querying, a method can then find and use the graph-based index corresponding the range of a query.
As the graph includes exactly all in-range data objects, RFANN query is reduced to an ANN query.
However, as explained already, the number of possible query ranges,  each in the format of $[L, R]$ with $1\le L \le R$, is $O(n^2)$. Therefore, compression is applied. However, the lossy compression reduces the quality of the indices, causing suboptimal query performance.

\begin{figure}[!htp]
 
    \centering
    \includegraphics[width=\linewidth]{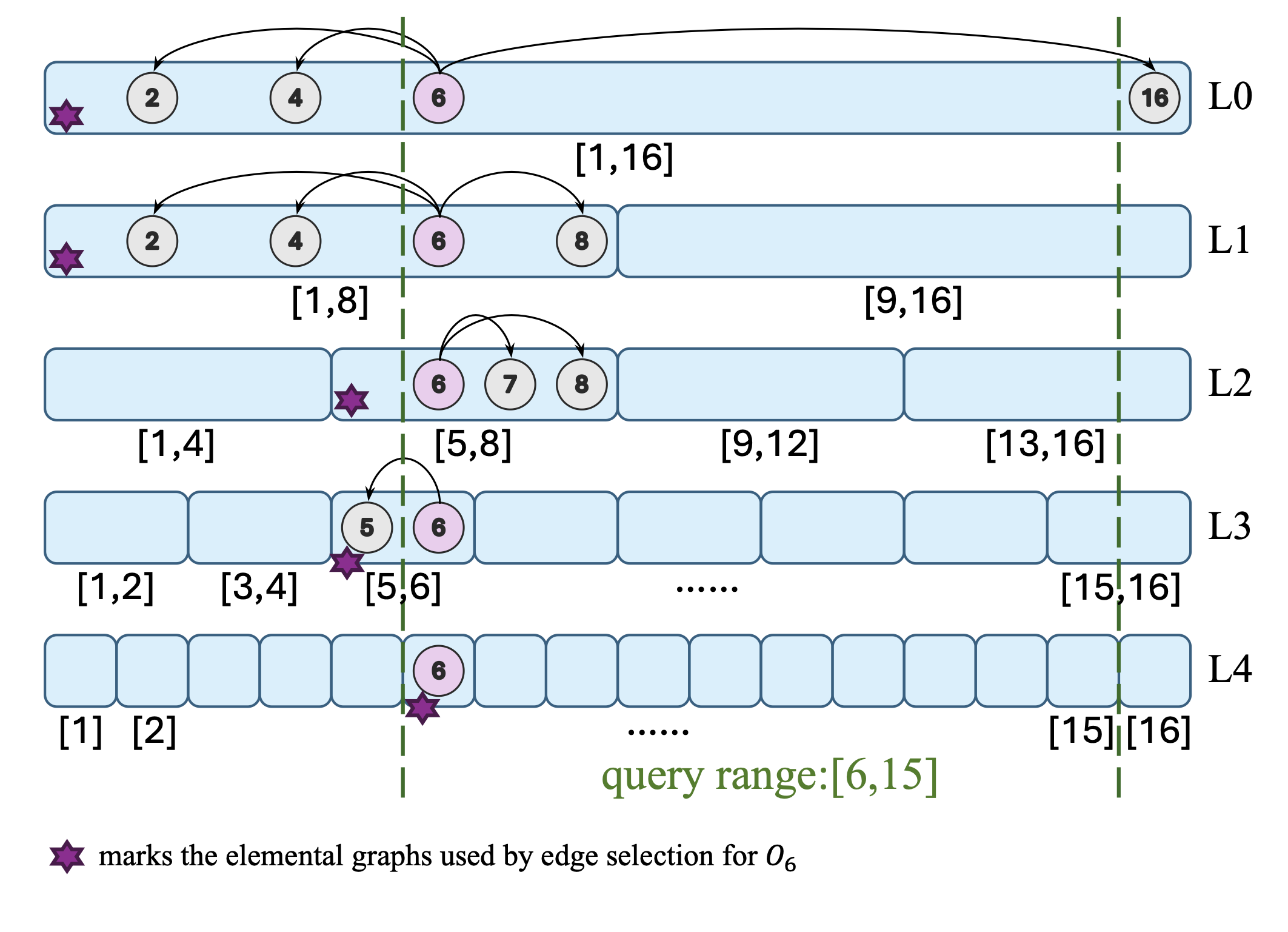}
   
    \vspace{-6mm}
    \caption{
    The \texttt{iRangeGraph} index applied to 16 data objects. It is based on a segment tree with 5 layers, namely L0 to L4. L0 has one segment corresponding to the range [1, 16]. L1 has two segments corresponding to the ranges [1, 8] and [9, 16] respectively, etc.
    The elemental graphs are materialized for each segment with respective data objects (e.g., an elemental graph based on $O_1,O_2,...,O_8$ is materialized for segment [1, 8] and the out-going edges of node $O_6$ are represented by the arrows).
    In all elemental graphs, the maximum out-degree $m$  of a node is 3 in this example.}
    \label{fig:STMG}
    \vspace{-2mm}
\end{figure}

\section{methodology}
\label{sec:methodology}

\subsection{Overview}
\label{subsec:overview}
We follow the idea of performing the search process on a dedicated graph-based index for a given query range as the existing method~\cite{segmentgraph} does.
However, while that method considers all $O(n^2)$ possible ranges and materializes compressed graphs for these ranges, we build graphs for only a moderate number of ranges before the querying.
We call these graphs \textit{elemental} graphs. Then when a query is processed, the edges in elemental graphs are used as the basic elements to construct a dedicated graph for the query's range.
The intuition is that a query range will intersect with some ranges for which elemental graphs exist.
The edges in these graphs are leveraged to construct the dedicated graph.
Dedicated graphs are constructed \emph{on-the-fly} in that we determine the edges of a data object only when we are to visit its neighbors during the search process.
The construction process incurs only a modest overhead compared to the cost of searching the dedicated graph.
Overall, the dedicated graph for a query range is not materialized but is constructed on-the-fly with little overhead.
This eliminates the huge space cost of storing all dedicated graphs but can still leverage a dedicated graph for any query range in the search process.
Two questions still need answers: (1) How to select the ranges for which elemental graphs are materialized? (2) How to construct a dedicated graph for any given query range based on elemental graphs?
In particular, the main consideration when answering these questions is the trade-off between a) the number of ranges for which elemental graphs should be built and b) the number of elemental graphs the algorithm needs to access when constructing the out-edges for a vertex on-the-fly for a given query range. 
The former determines the space complexity of the index, while the latter determines the time complexity of querying. 
We notice that the classical segment tree has promising properties related to the handling of range queries and strikes a good balance between these two costs, as illustrated next.

For the \underline{\textbf{first}} question, inspired by the classical segment tree~\cite{SegmentTree}, we build a multi-layer index structure (see Figure~\ref{fig:STMG}). In the $i$th layer ($i = 0, 1, ..., \log n$), the full range $[1,n]$ (corresponding to the set of all objects $O_1$, $O_2$, ..., $O_n$) is partitioned into $2^i$ disjoint segments (i.e., ranges), each of length $\frac{n}{2^i}$.
For each segment, we materialize an RNG-based elemental graph for the corresponding range~\footnote{ To ease the presentation, we assume that $n$ is power of 2.}.
The rationale for choosing the segments indexed by the segment tree and build graphs for them is three-fold. (1) The space consumption is moderate. In the multi-layer index, a data object appears only once in each layer, and it has a maximum out-degree of $m$ in a graph of each layer. Thus, the space complexity of storing the graphs for $n$ data objects is $O(nm \log n)$, which is much lower than the prohibitive space consumption of storing graphs for all possible query ranges, i.e., $O(n^3m)$. (2) These elemental graphs enable an effective and efficient algorithm for constructing a dedicated graph for any given query range  during querying (details are given in Section~\ref{subsec:query}).(3) The recursive structure of the segment tree enables an efficient algorithm for materializing the graphs in the index phase. It can be shown that the asymptotic time complexity of constructing the entire multi-layer index differs by up to a sub-logarithmic factor from that of constructing a single graph-based index (HNSW) for the dataset.
Section~\ref{subsec:index} elaborates the index phase.

To answer the \underline{\textbf{second}} question, we design an effective and efficient algorithm for constructing an RNG-based dedicated graph for any given query range on the fly.
In particular, we construct the graph by selecting the edges for an in-range data object from the elemental graphs when we need to visit its neighbors during the search process.  Each object is covered by a segment in our index at each of the $O(\log n)$ layers, and hence the object is involved in $O(\log n)$  elemental graphs. 
(1) The algorithm is effective at constructing an RNG-based dedicated graph because an RNG is defined by a pruning rule ( Section~\ref{subsec:anns_and_graph}), and the relationship between the query range and a certain segment enables determining whether an edge in an elemental graph of the segment should be retained in the dedicated graph for the given query range. 
For example, when a segment covers the given query range, the edges (connecting two in-range objects) that are retained in the elemental graph of the segment must be retained in the dedicated graph of the query range (since if an edge cannot be pruned from a set of objects, it can not be pruned for a subset of objects, either).
(2) Our algorithm is efficient because the index is based on the segment tree whose recursive structure allows us to skip the edge selection in some layers without affecting the dedicated graph.
Based on the idea of skipping layers, we design an algorithm for selecting edges for an object, which has amortized time complexity $O(m+\log n)$ - as a comparison, the time complexity is $O(m\log n)$ when enumerating the edges in $O(\log n)$ layers without skipping layers.
This is low compared to the cost of visiting the neighbors of an object in the dedicated graph, which is $O(md)$, where $d$ is the dimensionality of the vectors.
Section~\ref{subsec:query} presents the details of the algorithm for constructing a dedicated graph and the search strategies in the query phase.

 We summarize the resulting method, called \texttt{iRangeGraph} and provide an analysis of it in Section~\ref{subsec:summary}.

\subsection{Materializing Elemental Graphs}
\label{subsec:index}
\subsubsection{The Structure of the Index}
Our index is based on segment tree~\cite{SegmentTree}, which is a powerful data structure for supporting various range-based queries (e.g., range maximum query, range sum query, etc).  
Specifically, it is a balanced binary tree of $O(\log n)$ layers (see Figure~\ref{fig:STMG}). Each node in the tree represents a segment (i.e., a range), which is defined recursively.
In particular, the root node is defined to represent the full range of the dataset, i.e., $[1,n]$ (i.e., corresponding to the set of all objects $O_1,O_2,...,O_n$). Then recursively, for a node whose corresponding segment is $[l,r]$, its left and right children are defined to represent the segments of $\left[l, mid\right]$ and $\left[mid +1 , r\right]$ respectively where $mid = \lfloor \frac{l+r}{2} \rfloor$. 
Those nodes whose segments have the length of 1 are defined to be the leaf nodes (i.e., corresponding to a single object).

Based on the recursive structure of the segment tree, we note that a data object appears only once in a single segment in each of the $O(\log n)$ layers. 
Recall that a graph-based index has the maximum out-degree of $m$ for each data object.
Thus, when we build a graph for every segment in the segment tree, the space complexity of storing the index is $O(nm\log n)$, which is moderate compared with that of storing the graphs for all possible query ranges, i.e., $O(n^3 m)$. 
In other word, it is feasible to materialize all the graphs for the segments in the segment tree in the index phase. These materialized graphs 
which we call the \textit{elemental graphs}, will be later used in the query phase for constructing a dedicated graph for any given query range ( details will be presented in Section~\ref{subsec:query}).

For the graph indexes that are built for the segments (i.e., elemental graphs), we follow state-of-the-art graph-based algorithms~\cite{2019diskann,nsg,hnsw} and adopt approximate RNG-based graphs.
Recall that the time complexity of exactly computing an RNG is prohibitive~\cite{2019diskann,nsg}. 
We follow the convention of most of the existing methods~\cite{2019diskann,nsg,hnsw} which constructs approximate RNGs by searching for a few candidates, applying the pruning rule of RNG and cutting off excessive edges (see Section~\ref{subsec:anns_and_graph} for detailed description).

\subsubsection{The Algorithm for Materializing the Index}
\label{subsubsec: materialize elemental graph}
Next, to materialize the index, we note that it can be achieved trivially by running the construction algorithm of any existing graph indexes~\cite{2019diskann,nsg,hnsw} for every segment independently. 
The good news is that, due to the recursive structure of the segment tree, the construction can be done in a bottom-up manner, which is more efficient (e.g., it can be proven that the time complexity for constructing the entire index differs by up to a sub-logarithmic factor from that for constructing HNSW on the same dataset, details are given in Section~\ref{subsec:summary}), as follows.

Let $[l,r]$ be a segment in the tree and $[l,mid],[mid+1,r]$ be its left and right child segment respectively. 
Consider the procedure of constructing the approximate RNG of $[l,r]$, for which the graphs for $[l,mid]$ and $[mid+1,r]$ are available given that we construct the index in a bottom-up manner.
Recall that for constructing the edges of a data object $u$, we first find some candidates (e.g., approximate nearest neighbors) and then apply the pruning rule of RNG to them. 
Without loss of generality, assume that $u$ is covered by the child segment $[l, mid]$ (the case where $u$ is covered by the other child segment $[mid+1, r]$ is symmetric and thus omitted).
Note that the candidates are from either $[l,mid]$ (the child segment that contains $u$) or $[mid+1,r]$ (the child segment that does not contsain $u$).
We discuss these two cases separately.

\underline{First}, for the candidates that are from the child segment that contains $u$, i.e., $[l,mid]$, we can infer whether a candidate of object $u$ will be pruned in the RNG of $[l,r]$ based on the RNG of its child segment $[l,mid]$. 
In particular, we note that a candidate may not be pruned in the RNG of $[l,r]$ only if it is retained in the RNG of the child segment $[l,mid]$. 
Note that the set of objects of the child segment $[l,mid]$ is a subset of that of the segment $[l,r]$. 
Thus, if a candidate can be pruned by an object in the subset, it can also be pruned by the same object in the full set. 
Thus, for generating the candidates that are from $[l,mid]$, it suffices to consider those that cannot be pruned in the RNG of the child segment only (i.e., to copy the neighbors of $u$ in the elemental graph of $[l,mid]$ that have been constructed). 
It is unnecessary to consider other objects in $[l,mid]$ as the candidates since otherwise they would be pruned. 
This operation avoids the cost of searching for candidates of $u$ in $[l,mid]$ from scratch.
\underline{Second}, for the candidates that are from the child segment that does not contain $u$, i.e., $[mid+1,r]$, we must search for candidates for $u$ as we have no clue which candidates from $[mid+1,r]$ must be pruned in this case. 
In particular, in this case, we take the approximate nearest neighbors as the candidates by following the well-known HNSW~\cite{hnsw} method.

After the candidates are generated, finally the pruning of RNG is applied to the candidates, and the edges are built accordingly. 
In summary, in the index phase, we materialize the RNG-based elemental graphs for all the segments in the segment tree bottom-up, which is conducted recursively.

\subsection{Constructing and Searching Dedicated Graphs}
\label{subsec:query}

\subsubsection{ Constructing Dedicated Graphs}
\label{subsubsec: edge selection}
Let $[L,R]$ be a query range given in the query phase. Recall that we target to construct a dedicated RNG-based graph for any given query range on the fly for the RFANN query. Basically, the construction of a graph is to provide the set of edges for every data object $u \in [L,R]$. 
In our index, note that $u$ is included in a single segment in each of the $O(\log n)$ layers (see Figure~\ref{fig:STMG}). As in the elemental graph of each segment, a data object has up to $m$ edges, a data object has in total $O(m\log n)$ edges in the index. 
To construct the dedicated graph for the query range, the question is how we should select up to $m$ edges out of the $O(m\log n)$ edges in the elemental graphs.
For example, in Figure~\ref{fig:STMG}, the object $O_6$ has 9 edges in total from all layers, and in order to construct the dedicated graph, we need to select among them up to 3 edges for $O_6$.

\smallskip\noindent\textbf{ The Strategy of Edge Selection.}
The relationship between a segment and the query range can help to decide whether an edge in the elemental graph should be selected in the dedicated graph for the query range.
For example, the segments that contain an object $u$ (in Figure~\ref{fig:STMG}, the purple stars mark all the segments that contain data object $O_6$) all intersect with the query range, and the larger the intersection is, the more strict the pruning would be (meaning that it is more likely that an edge is pruned).
Specifically, if an edge $(u,v)$ is retained in the elemental graph of the segment $[l,r]$, this indicates that the edge would not be pruned by any of the objects in the subset $[L,R]\cap [l,r]$ of the query range $[L,R]$. 
In the extreme case, where $[L,R]\cap [l,r]=[L,R]$, i.e., the segment covers the query range, it is guaranteed that the edge would not be pruned by any of the objects in the query range.
In other word, if an edge is retained in the elemental graph that has larger intersection with the query range, it is more robust against pruning by the objects in the given query range.
As a result, given that we only select $m$ edges out of the $O(m\log n)$ edges, we give priority to selecting the edges in the segment that has larger intersection with the query range. In our index, they correspond to the edges in the upper layers of the segment tree.

\smallskip\noindent\textbf{ The Efficient Algorithm for Edge Selection.}
Based on the above strategy of edge selection, for a data object $u$, we select the edges in a top-down way. 
Specifically, we start the process from the root node of the segment tree. Then iteratively, we select all the in-range edges $(u,v)$ in the  elemental graph of a layer and move to the segment that contains $u$ in the next layer. 
The edge selection terminates after selecting edges of an elemental graph whose segment is covered by the query range.
This is because if an edge can be pruned by an object in this segment, then as the object is also in the query range, the edge is also pruned in the RNG-based graph on the query range by the same object.

Although this algorithm is effective in constructing a dedicated graph for a given query range, its time cost in the query phase is unignorable. 
Specifically, in the worst-case, the cost of selecting the edges for an object is $O(m\log n)$: it scans $m$ edges of the object at each of the $O(\log n)$ segments covering the object and thus takes $O(m \log n)$ time.
For example, for the query range [6, 15] and data object $O_6$ shown in Figure~\ref{fig:STMG}, in order to select up to 3 edges, it scans the edges of $O_6$ in every layer, i.e., it scans 9 edges at 4 layers in total.
Recall that our strategy is to prioritize the edge selection in the segment that has the larger intersection with the query range.
In particular, when a segment has exactly the same intersection with the query range as its child segment does, the edges in their elemental graphs have exactly the same robustness against pruning by the objects in the query range. 
In this case, we can skip the edge selection at this layer and directly move to the next layer.
For example, in Figure~\ref{fig:STMG}, 
to select the edges for $O_6$, the segments in L1 and L2 that contain $O_6$ have exactly the same intersection with the query range, and thus we can skip L1 to select edges from L2.

We present our edge selection algorithm based on the strategy of skipping layers in Algorithm~\ref{alg:edge selection}.
We note that it is provable that the amortized time complexity of the algorithm is $O(m + \log n)$, which is much smaller than the time complexity of an edge selection algorithm that does not skip layers (i.e., $O(m \log n)$).
In addition, the time complexity of our edge selection algorithm is negligible compared with the time cost of visiting the $m$ neighbors, which is $O(md)$, where $d$ is the dimensionality (see details in Section~\ref{subsec:summary}).

\subsubsection{ Searching on Dedicated Graphs}
For a given RFANN query, we search on the dedicated graph, which we construct on the fly for the query range, 
as we do on existing graph-based indexes for the ANN query. Specifically, we employ greedy beam search on the dedicated graph, where the edges for a data object are constructed on the fly when we 
access its neighbors during the search process.

\begin{algorithm}[!htbp]
\caption{Edge Selection}
\label{alg:edge selection}
\begin{algorithmic}[1] 

\Require{$u$: a data object; $[L, R]$: the query range; $m$: the maximum out-degree;  $\mathcal{N}_{\mathit{lay},u}$: the set of neighbors of $u$ at layer $\mathit{lay}$}
\Ensure{a set of neighbors selected for $u$}

\State $l\gets 1, r\gets n,  \mathit{lay} \gets 0, \mathcal{S}\gets \emptyset$ 

\While{$|\mathcal{S}|<m$}
    \State $[l_{c},r_{c}] \gets$ the child segment of $[l,r]$ which contains $u$
    \If {$[l_{c},r_{c}] \cap [L, R] = [l,r] \cap [L, R]$} 
        \State $l \gets l_{c}, r \gets r_{c}, \mathit{lay} \gets \mathit{lay} + 1$
    \Else 
    \State $\mathcal{S} \gets \mathcal{S} \cup (\mathcal{N}_{ \mathit{lay},u}\cap [L,R])$
    \State Cut off $\mathcal{S}$ to the size of $m$
    \If {$[l,r] \subseteq [L, R]$} 
        \State \textbf{break}
    \EndIf
    \State $l \gets l_{c}, r \gets r_{c},  \mathit{lay} \gets \mathit{lay} + 1$

    \EndIf
\EndWhile

\State \Return{$\mathcal{S}$}

\end{algorithmic}
\end{algorithm}

\subsection{Summary and Discussion}
\label{subsec:summary}

In summary, the proposed method initially builds a multi-layer index based on the segment tree, where for each segment, we materialize an approximate RNG-based graphs, called an elemental graph.
During query processing, the method leverages the elemental graphs to construct a dedicated graph for the query being processed on the fly (Algorithm~\ref{alg:edge selection}) and conduct an ANN query on it.

\smallskip\noindent\textbf{Analysis.} 
First, \textbf{the space complexity} of our method is of $O(nm\log n)$ which is moderate compared with that of storing the dedicated graphs for all possible query ranges, i.e., $O(n^3 m)$. 
Next, \textbf{the time complexity of the index phase} is a somewhat controversial topic. This is because the construction algorithm involves nearest neighbor search.
It can be proven that no algorithm can be guaranteed to achieve sub-linear time complexity for nearest neighbor search~\cite{hardness_ANN, indyk1998approximate} (including graph-based methods~\cite{indyk2023worstcase}). However, due to the promising performance of the graph-based methods in practice on real-world datasets, it is often the case that these methods have sub-linear performance on real-world datasets~\cite{hnsw,nsg,2019diskann, graph_benchmark}. 
Thus, instead of presenting an asymptotic time complexity for the construction algorithm, we argue that the time complexity of constructing the entire multi-layer index differs by up to a sub-logarithmic factor from that of constructing a single graph-based index (e.g., HNSW) for the same dataset. 
The conclusion holds regardless of the above issue (see Theorem~\ref{theorem:construction time complexity} below). 
Empirically, the time cost for constructing the entire index of our method is no more than 3x of that for constructing HNSW over the set of all objects (note that HNSW only supports queries without range constraints).
The detailed proof and empirical verification are in the technical report~\cite{technical_report} due to the page limitation.

\begin{theorem}
    \label{theorem:construction time complexity}
    The time complexity of the algorithm for materializing the entire index of our method differs by up to a sub-logarithmic factor from that for constructing HNSW on the set of all objects.
\end{theorem}

As for \textbf{the time complexity of querying}, due to the same issue as above, it is not informative to compare two graph-based methods based on their big-O time complexity (because for all methods, the time complexity is $O(n)$). We note that for the RFANN query, our method is superior because (1) it constructs a dedicated RNG-based graph for any given query range in the query phase, and (2) the time cost of doing so is low (see Theorem~\ref{theorem:edge selection time complexity} below, the detailed proof is given in the technical report~\cite{technical_report}).
We refer to the empirical study in Section~\ref{subsec:exp single attribute} that provides evidence that our method is capable of state-of-the-art performance.

\begin{theorem}
    \label{theorem:edge selection time complexity}
    The amortized time complexity of the algorithm (Algorithm~\ref{alg:edge selection}) for constructing the edges for a data object in the dedicated graph on the fly is $O(m+\log n)$.
\end{theorem}

\smallskip\noindent \textbf{Limitation.} 
Nevertheless, we note that the dedicated graph constructed by our algorithm might not approximate an RNG perfectly based on all in-range objects for a query range, i.e., the graph constructed using elemental graphs is not necessarily identical to the corresponding dedicated graph built from scratch.
The space complexity of storing the graphs that are explicitly materialized on the in-range objects for all possible query ranges is $O(n^3m)$, while the space complexity of our method is only $O(nm\log n)$. 
We observe that the elemental graphs may miss some edges that should have been retained in the approximate RNG of the query range. 
Specifically, it is possible that an edge that should be retained in the RNG of the query range is pruned in the elemental graph of a segment. This is because besides the in-range objects, a segment also involves some out-of-ranges objects, which may accidentally prune the edges.
Despite this limitation, our method achieves search performance that is very close to the ideal search performance, as verified empirically in Section~\ref{subsec:exp oracle hnsw}. Specifically, in Section~\ref{subsec:exp oracle hnsw}, we explicitly materialize an HNSW for each query range in a query workload. 
These HNSWs are called Oracle-HNSW because materializing them for all possible query ranges is impractical.
We measure the search performance of our method and the Oracle-HNSW, finding that the queries per second of the impractical Oracle-HNSW is no more than 2x that of ours when reaching 0.9 recall.
At the same time, the space complexity of our method is $O(nm\log n)$, which is much smaller than that of the Oracle-HNSW, which is $O(n^3m)$.

\smallskip\noindent \textbf{Comparison with \texttt{SuperPostfiltering}~\cite{WindowFilter}.} 
We notice that a concurrent study~\cite{WindowFilter}, which was recently posted on arXiv, proposes a method called \texttt{SuperPostfiltering} for the RFANN query with a method also inspired by the segment tree. 
Unlike our method that aims to build a dedicated graph for any given query range on the fly, \texttt{SuperPostfiltering}~\cite{WindowFilter} builds graph-based indexes for multiple ranges in the index phase. In the query phase, it then adopts the corresponding graph of the smallest range that covers the query range for conducting \texttt{Post-filtering} RFANN search.
As a result, this method still suffers from the inherent issues of \texttt{Post-filtering} (see Section~\ref{subsec:RFANNS}), i.e., it may visit many out-of-range objects for finding the nearest in-range object.
Furthermore, the concurrent study~\cite{WindowFilter} only applies the segment tree for determining the ranges to build graphs for. 
In our method, the segment tree is applied in a more integrated way, i.e., the recursive structure of the segment tree helps both the efficient construction of our index structure in the index phase (Section~\ref{subsec:index}) and the effective and efficient strategy of edge selection for constructing dedicated graphs on the fly in the query phase (Section~\ref{subsec:query}).
According to the empirical study in Section~\ref{subsec:exp single attribute}, our method is in general superior to \texttt{SuperPostfiltering} in terms of search performance, memory footprint, and indexing time.

\smallskip\noindent \textbf{Impact of Duplicate Attribute values and Cardinality of Attribute.}  
So far, we have assumed that all data objects have distinct attribute values. 
However, it is easy to extend the algorithm to handle duplicate attribute values.
Specifically, we can simply put the data objects with identical attribute values into the same node in every layer of the segment tree. All other operations in the algorithm remain unchanged.
Furthermore, the RFANN query becomes easier when duplicate attribute values exists. 
Specifically, when the number $c$ of distinct attribute values, i.e., the cardinality of the attribute, is significantly smaller than $n$, the space complexity of our index becomes $O(nm\log c)$, and the time complexity in Theorem~\ref{theorem:edge selection time complexity} becomes $O(m + \log c)$.
On the other hand, we note that even when $c$ is small, the cost of storing a graph-based index for every possible query range remains impractical. This is because the space complexity of storing indexes for all possible query ranges is $O(nmc^2)$. For example, when $c=100$, 5,050 indexes must be stored, which incurs impractical costs.

\section{Multi-Attribute Range-Filtering ANN Search}
\label{sec:MultiAttribute}
We have developed an algorithm for the RFANN that involves a predicate on a single numeric attribute.
In real-world scenarios, queries that involve conjunctive predicates~\footnote{It refers to the queries that have multiple constraints on multiple attributes. A data object is said to satisfy the conjunctive predicates if and only if it satisfies all the constraints.} on multiple numeric attributes are also common~\cite{wang2021milvus,vbase,HQI}. 
For example, in a passage retrieval system, clients could set constraints on publish time and passage length in order to find desired results.
In particular, to support multi-attribute queries, during the index phase, we can build our index with respect to one of the numeric attributes, e.g., $A_1$.
Then, during the query phase, we can construct a dedicated graph-based index for the query range on $A_1$.
Then for a multi-attribute query, we can conduct the graph-based ANN search with the dedicated graph built based on $A_1$ and adopt the \texttt{In-filtering} or \texttt{Post-filtering} strategy for handling the predicates on other attributes (see Section~\ref{subsec:RFANNS}).

While with this simple extension, our method can support multi-attribute queries, we note that due to the inherent issues of the basic strategies, the extended algorithm may suffer from degenerated performance for some workloads.
For example, as discussed in Section~\ref{subsec:RFANNS}, the \texttt{In-filtering} strategy may have the issue that a data object has few (or no) in-range neighbors. 
On the other hand, the \texttt{Post-filtering} strategy may have the issue that it may visit a large number of out-of-range objects to find an in-range object.
To mitigate these issues, we propose a simple idea that generalizes the \texttt{In-filtering} and \texttt{Post-filtering} strategies.
Note that the data objects in the dedicated graph-based index are all in-range in terms of $A_1$. Next, by ``in-range'' and ``out-of-range'', we refer to the ``in-range'' and ``out-of-range'' objects w.r.t. the attributes other than $A_1$.
Specifically, when conducting graph-based ANN search, unlike the \texttt{In-filtering} (resp. \texttt{Post-filtering}) strategy that visits none (resp. all) of the out-of-range neighbors, we visit the out-of-range neighbors with probability of $p$ where $0\le p\le 1$.
When $p=0$, the method is exactly equivalent to the \texttt{In-filtering} strategy, i.e., it does not visit the out-of-range objects. 
When $p=1$, the method is equivalent to the \texttt{Post-filtering} strategy, i.e., it visits all the neighbors regardless of their attribute values during the graph-based ANN search and finds the nearest neighbors from only the in-range objects that are visited.
Thus, when $0<p<1$, it corresponds to an intermediate strategy in-between \texttt{In-filtering} and \texttt{Post-filtering}, which may mitigate the disadvantages of the two extreme strategies.
Our experimental results in Section~\ref{subsec:exp multi attribute} indicate that this simple technique is capable of a $70\%$ speed-up at 0.9 recall.

\section{experimental study}
\label{sec:experiment}

\subsection{Experimental Setup}
\label{subsec:experimental setup}
Our experiments involve four studies. 
(1) We compare our method with the existing methods for the single-attribute RFANN query (Section~\ref{subsec:exp single attribute}). 
(2) We verify the effectiveness of the proposed techniques via ablation studies (Section~\ref{subsec:exp ablation study}) and the scalability study (Section~\ref{subsec:exp scalability}). 
(3) Recall that the target of our method is to build a dedicated graph for any given query range. We measure the performance gap between our method and the dedicated graph which is explicitly 
materialized for the given query ranges (Section~\ref{subsec:exp oracle hnsw}).
(4) We evaluate the extension of our method on the multi-attribute RFANN query (Section~\ref{subsec:exp multi attribute}). 
In all of the experiments, the RFANN query targets to find the first $10$ nearest neighbors that satisfy the predicates.

\smallskip\noindent\textbf{Datasets.}
For evaluating the performance of the methods, We adopt five real-world public datasets including WIT-Image~\footnote{\url{https://github.com/google-research-datasets/wit}} (WIT in short), TripClick~\footnote{\url{https://tripdatabase.github.io/tripclick/}}, Redcaps~\footnote{\url{https://redcaps.xyz/}}, YouTube-Rgb~\footnote{\url{https://research.google.com/youtube8m/download.html}} (YT-Rgb in short) and YouTube-Audio~\footnote{\url{https://research.google.com/youtube8m/download.html}} (YT-Audio in short).
These datasets have been used in existing studies on RFANN~\cite{segmentgraph,WindowFilter,patel2024acorn}.
For each dataset, one million objects which involve both real-world vectors and numeric attributes are extracted to be the data objects. Another 1,000 vectors are extracted to be the query vectors. For these query vectors, we would assign different query ranges to them in order to test the performance of the methods for different workloads. The details will be specified in the next paragraph. 
The properties of the datasets are summarized in Table~\ref{tab:datasets}. 
Note that YT-Rgb and YT-Audio involve two attributes. We use the first attribute (i.e., \# of likes for YT-Rgb and publish time for YT-Audio) for the evaluation of the single-attribute RFANN query and use both for the evaluation of the multi-attribute RFANN query.
The detailed descriptions of the datasets are left in the technical report~\cite{technical_report} due to the page limit.

\begin{table}[!htbp]
    \centering
    \vspace{-2mm}
    \caption{Datasets. All the datasets involve one million data objects and one thousand queries.}
    \vspace{-4mm}
    \begin{tabular}{c|c|c|c}
    \toprule
          & Vector Type & Dim. & Attribute Type\\ \hline

     WIT  & image & 2,048 &  image size \\
     TripClick   &  text & 768 & 
      publication date
     \\
     Redcaps  & multi-modality & 512 &  
      timestamp
     \\
     YT-Rgb  & video & 1,024 &  \# of likes, \# of comments \\
     YT-Audio & audio & 128 &  publish time,  \# of views \\
     
    \bottomrule
    \end{tabular}
    \label{tab:datasets}
\end{table}

\smallskip\noindent\textbf{Query Ranges.}
We follow several studies~\cite{HQI,wang2021milvus,segmentgraph} by adopting datasets that feature real-world attributes and evaluating query performance across varying query workloads. 
We note that although it is mentioned in many industrial studies that the RFANN query is an important feature in real-world systems (e.g., Apple~\cite{HQI}, Milvus~\cite{wang2021milvus}, Alibaba~\cite{wei2020analyticdb}). However, possibly due to privacy issues, no publicly available datasets with range-filters have been provided from industry, to the best of our knowledge.
Instead, we evaluate the search performance of the methods under different query ranges by following  the existing study~\cite{HQI}. 
In particular, we say a query has the {\textit{range fraction}} of $2^{-i}$ if its query range covers $n/2^i$ of the data objects.
Based on the range fractions, the queries can be divided into three scales including large ($i\in[0,3]$), moderate ($i\in[4,6]$), and small ($i\in[7,9]$). 
As for even smaller query ranges, we note that these cases are inherently simple for the RFANN query. They can be handled efficiently with the simple \texttt{Pre-filtering} strategy. Thus, we exclude them from the evaluation.
We evaluate the methods in two types of workload including (1) the workload with a fixed range fraction and (2) the workload with mixed range fractions. 
For the fixed workload, when the length of the query range (i.e., the range fraction) is fixed, the specific locations of the query ranges are generated randomly for the query vectors.
For the mixed workload, we first randomly partition the query vectors into 10 subsets. 
For the $i$th subset ($i\in [0,9]$), we assign a query range with the range fraction of $2^{-i}$ to the queries
and generate the specific query ranges randomly.

\smallskip\noindent\textbf{Performance Metrics.}
Following existing benchmarks~\cite{annbenchmark,graph_benchmark}, we primarily measure the efficiency by qps, i.e., the number of queries responded per second, and we measure the accuracy by $recall=\frac{|G\cap S|}{K}$, where $G$ is the groundtruth of the $K$ nearest neighbors, and $S$ is the results produced by an algorithm.
In addition to qps and recall, we note that there are other metrics which provide valuable insight into the evaluation of ANN algorithms~\cite{patella2008many}, e.g., the number of distance computations and average distance ratio. 
For brevity, we include experimental results regarding these metrics in the technical report~\cite{technical_report}.

\smallskip\noindent\textbf{Methods and Parameters.}
We study 6 methods, including our method and 5 existing methods as follows.
For the methods which provide default parameters, we adopt the default parameters; for others we use grid search to find the optimal parameters for building the indexes.
We vary the parameter named the beam size for all the graph-based methods (i.e., except \texttt{Pre-filtering}) to control the qps-recall trade-off in the query phase.
\underline{ (1)\texttt{~iRangeGraph} } is our method. For index construction, there are two parameters, maximum out-degree $m$, and the number of candidates generated for constructing the graphs $EF$ (see Section~\ref{subsec:index}).
$m$ is set to $16$ for TripClick, YT-Audio and $64$ for WIT, Redcaps and YT-Rgb. $EF$ is set to $100$ for WIT, TripClick and YT-Audio and $400$ for Redcaps and YT-Rgb. \underline{(2)\texttt{~2DSegmentGraph}~\cite{segmentgraph}} builds compressed dedicated graphs for all possible query ranges.
In particular, it provides three versions of index construction named MaxLeap, MidLeap and MinLeap. As reported in the paper, the MaxLeap version shows optimal search performance with significantly smaller index time and space.
Therefore, we evaluate \texttt{2DSegmentGraph} with MaxLeap.
We follow the parameter setting in~\cite{segmentgraph} for WIT and YT-Audio, i.e., $M=8$, $K=100$ for YT-Audio, and $M=64$, $K=100$ for WIT, where $M$ denotes the maximum out-degree and $K$ is a parameter which controls the index construction. We set $K=100$, $M=32$ for all the rest datasets based on grid search.
\underline{(3)\texttt{~Filtered-DiskANN}~\cite{filtereddiskann}} proposes two methods for index construction, namely
\texttt{FilteredVamana} and \texttt{StitchedVamana}. 
We note that they are originally designed for label filtering instead of range filtering. 
Following the prior work~\cite{segmentgraph}, to adapt them to RFANN query, we evenly divide the full range $[1,n]$ into 10 consecutive buckets and assign each bucket a label. 
The buckets which have overlap with the query range are used as the query labels. We use the default parameters~\cite{github-filtereddiskann}, i.e., $R=64$, $L=100$ for both \texttt{FilteredVamana} and \texttt{StitchedVamana}, and set $SR=64$ for \texttt{StitchedVamana}.
\underline{(4)\texttt{~Milvus}~\cite{wang2021milvus}} is a vector database system that supports range filtering. We choose HNSW as its index. With grid search, we fix $EF=400$ for all datasets, and set $M=16$ for YT-Audio, $M=32$ for Redcaps, TripClick and YT-Rgb, and $M=64$ for WIT.
\underline {(5)\texttt{~SuperPostFiltering}~\cite{WindowFilter}} is a method based on post-filtering. The detailed discussion about the method can be found in Section~\ref{subsec:RFANNS} and Section~\ref{subsec:summary}. 
We use its recommended parameters, i.e., $\beta =2,EF=500,m=64$ for all datasets.
\underline{(6)\texttt{~Pre-filtering}} conducts a linear scan with objects satisfying the range constraint.
Specifically, it first applies binary search to eliminate the out-of-range objects. Then, among the remaining objects, it computes the distances between their vectors and the query vector to find the NN. The time cost of the first step is almost negligible, while that of the second step is high as it computes many distances between high-dimensional vectors.
No parameter tuning is needed.
\texttt{VBASE}~\cite{vbase} is excluded from the comparison because it has been reported that it, as a system for generic attribute-filtering ANN queries has suboptimal search performance for the RFANN query~\cite{WindowFilter}.

\smallskip\noindent\textbf{Platform.} All experiments are conducted on a server with Intel(R) Xeon(R) Gold 6418H CPU@4GHz and 1TB of RAM under Ubuntu 22.04.4 LTS. 
All methods are implemented in C++ and are compiled using GCC 11.4.0 with \texttt{-O3 -march=native}. The search performance is evaluated using a single thread, and the indexing time is measured using 32 threads.
The source code is available at \url{https://github.com/YuexuanXu7/iRangeGraph}.

\subsection{Experimental Results}
\subsubsection{The Results of the Single-Attribute RFANN Query}
\label{subsec:exp single attribute}
This section compares our \texttt{iRangeGraph} method with the baseline approaches for single-attribute RFANN query.

\begin{figure*}[thbp]
\centering

\begin{subfigure}[b]{1\textwidth}
\includegraphics[width=\textwidth]{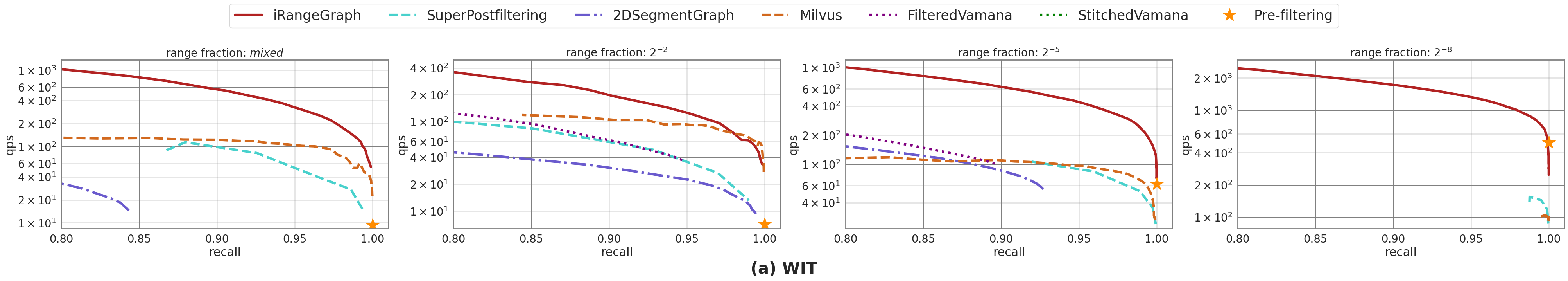}
\label{fig:main-wit}
\end{subfigure}

\vspace{-3mm}

\begin{subfigure}[b]{1\textwidth}
\includegraphics[width=\textwidth]{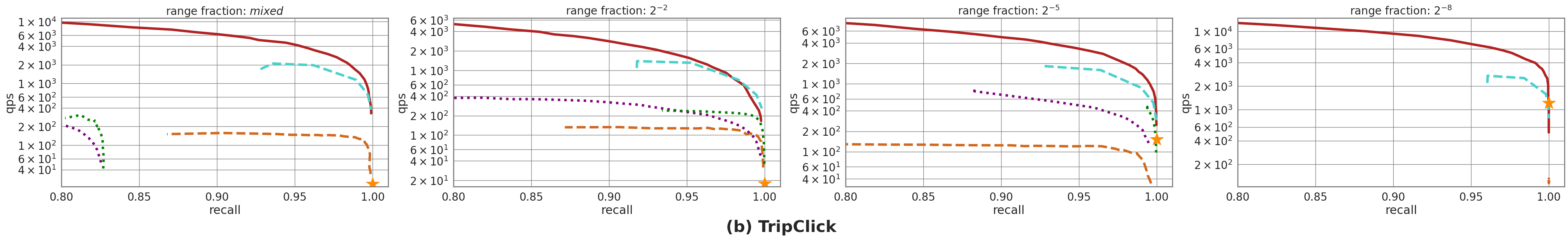}
\label{fig:main-tripclick}
\end{subfigure}

\vspace{-3mm}

\begin{subfigure}[b]{1\textwidth}
\includegraphics[width=\textwidth]{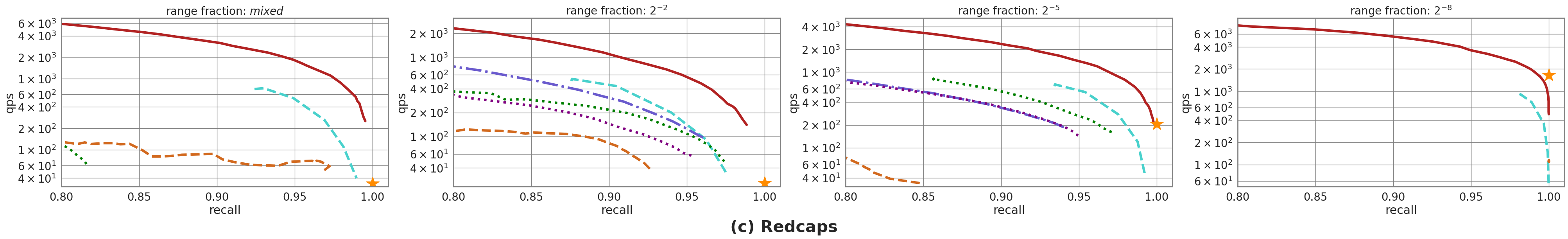}
\label{fig:main-redcaps}
\end{subfigure}

\vspace{-3mm}

\begin{subfigure}[b]{1\textwidth}
\includegraphics[width=\textwidth]{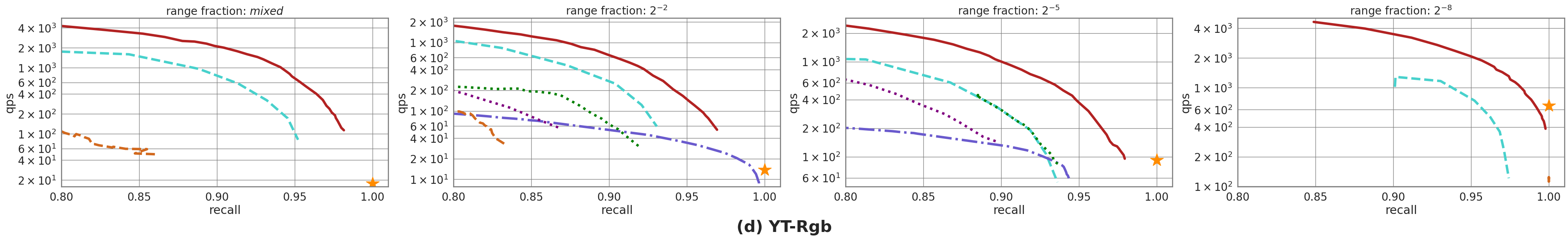}
\label{fig:main-rgb}
\end{subfigure}

\vspace{-3mm}

\begin{subfigure}[b]{1\textwidth}
\includegraphics[width=\textwidth]{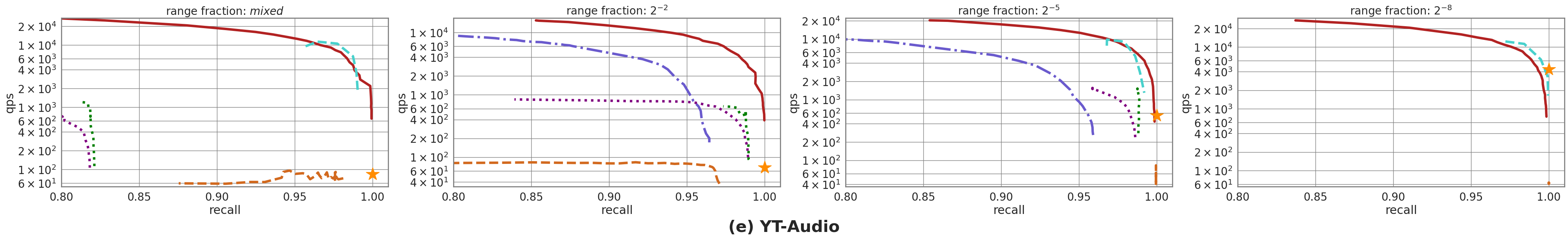}
\label{fig:main-audio}
\end{subfigure}

\vspace{-4mm}

\caption{Comparison of all methods on the single-attribute RFANN query with different datasets and query workloads of mixed, large, moderate and small range fractions. The curve of a method is missing for a certain dataset and query workload indicates that it fails to achieve at least 0.8 recall.}
\label{fig:main}
\vspace{-2mm}
\end{figure*}

\smallskip\noindent\textbf{The Search Performance for Single-Attribute RFANN Query.}
We first evaluate the search performance of all methods on RFANN queries. 
In particular, the search performance is evaluated under four different query workloads which correspond to the query ranges with different range fractions, including $2^{-2}$ (large scale), $2^{-5}$ (moderate scale), $2^{-8}$ (small scale) and the mixed.
Figure \ref{fig:main} plots the qps-recall curves (upper-right is better) by varying the parameter named beam size for the graph-based methods (i.e., the methods except for \texttt{Pre-filtering}). We have the following observations.
(1) Only our \texttt{iRangeGraph} method, \texttt{Pre-filtering}, \texttt{SuperPostFiltering} and \texttt{Milvus} could stably achieve reasonable recall in almost all datasets under the query workload of all the fixed and mixed range fractions. 
Other methods fail to achieve over 0.8 recall in the workload of the mixed range fractions in most of the datasets.
Note that in real-world scenarios, it is likely that a query workload involves query ranges with varying lengths.
(2) \texttt{2DSegmentGraph}, \texttt{FilteredVamana} and \texttt{StitchedVamana} can only handle the query workload of large and moderate range fractions while failing to achieve 0.8 recall on the query workload of small and mixed range fractions.
\texttt{2DSegmentGraph} may also fail in large and moderate range fractions on some datasets (e.g., on TripClick).
Note that the phenomenon that \texttt{2DSegmentGraph} cannot produce reasonable recall for small query ranges is consistent with the results in its paper~\cite{segmentgraph}.
The reduced query performance of \texttt{2DSegmentGraph} may be, to some extent, caused by its aggressive compression of dedicated graphs. 
In particular, based on the analysis of its compression strategy~\cite{segmentgraph}, the performance is only guaranteed on the queries with \textit{half-bounded} query ranges (i.e., $L=1$ or $R=n$). For other general ranges, the compression strategy is likely to cause reduced performance.
(3) In general, \texttt{iRangeGraph} shows clear superiority in the search performance over all baseline methods. 
Compared with the baselines, \texttt{iRangeGraph} achieves the best qps-recall trade-off in almost all datasets and query workloads. In particular, \texttt{iRangeGraph} surpasses the most competitive baseline \texttt{SuperPostfiltering} by 2x-5x in qps at 0.9 recall on most of the datasets (including WIT, TripClick, Redcaps, and YT-Rgb), and achieves even more improvement compared to other baselines, (e.g., 2x-300x speed-up over \texttt{Milvus}). 
On YT-Audio, it has comparable search performance with the most competitive baseline \texttt{SuperPostfiltering}.
This might be due to the dimensionality of the dataset, i.e., it has 128 dimensions only. In particular, when the dimensionality is small, the overhead of constructing the dedicated graph on the fly could somewhat affect the search performance, although the overhead has been reduced from $O(m\log n)$ to $O(m+\log n)$.
(4) For small range fractions, \texttt{Pre-Filtering} has the optimal efficiency if the target recall is 1. However, we note that our method still provides a better time-accuracy trade-off when the requirement on the recall is relaxed. For example, \texttt{iRangeGraph} achieves 3x-110x speedup over \texttt{Pre-filtering} at 0.9 recall among different datasets. 
Additional experimental results and analyses related to the single-attribute RFANN query are placed in the technical report~\cite{technical_report} due to the page limit.

\smallskip\noindent\textbf{The Memory Footprint and Indexing Time.}
We then measure the memory footprint and indexing time of all methods. We report the memory footprint of all the methods for RFANN in Table~\ref{tab:index size}. 
We also include the sizes of the raw vectors for reference. The overall memory footprint of a method minus that of the raw vectors equals to the size of its index.
The indexing time is reported in Table~\ref{tab:index time}~\footnote{ For all methods except for \texttt{2DSegmentGraph}, 32 threads are used; for \texttt{2DSegmentGraph}, a single thread is used since its multi-thread implementation is not available.}.
Note that only our method, \texttt{SuperPostfiltering}, \texttt{Milvus}, and \texttt{Pre-filtering} can stably produce reasonable recall for most of the datasets and query workloads. Thus, we focus the comparisons on these methods only. 
We note that the memory footprint of our method is smaller than that of \texttt{SuperPostfiltering} and larger than those of \texttt{Pre-filtering} and \texttt{Milvus}.
Recall that our method outperforms \texttt{Milvus} and \texttt{Pre-filtering} by orders of magnitudes in terms of query efficiency across many datasets and query workloads.
Thus, compared with these methods, our method strikes a good trade-off between the search performance and the (time and space) costs of the index.
On the other hand, as is reflected in Table~\ref{tab:index size}, it is not guaranteed that \texttt{iRangeGraph} has smaller memory footprint than \texttt{2DSegmentGraph} although \texttt{iRangeGraph} has the lower space complexity of $O(nm\log n)$. 
This is because \texttt{2DSegmentGraph} applies an aggressive compression strategy (MaxLeap) that produces an index whose empirical size is significantly smaller than $O(n^3m)$. 
However, due to the compression, \texttt{2DSegmentGraph} cannot enable reasonable query performance on many datasets and query workloads.
In contrast, our method achieves superior query performance with a stable and moderate memory footprint.

Based on the experimental results above, we proceed to characterize the settings in which deployment of
\texttt{iRangeGraph} is appropriate.
Overall, \texttt{iRangeGraph} is suited in setting where the query workload includes mixed query ranges and the dimensionality is high. 
In particular, \texttt{iRangeGraph} offers competitive query performance at the cost of moderate (but unignorable) memory consumption. Its memory overhead is caused by its storage of the edges of elemental graphs. When the dimensionality is high, the memory cost is dominated by the cost of storing the vectors. Thus, this overhead is relatively low.
For example, the WIT dataset has 2,048 dimensions, and the memory consumption of our method is ~1.6x that of the raw vectors.
As for the query performance, unless the query ranges are consistently small (in this case, the RFANN query is intrinsically simple and \texttt{Pre-filtering} is the optimal solution), \texttt{iRangeGraph} shows competitive search performance. 
We note that the methods designed for label filtering do not quite suit the RFANN query. Recall that these methods partition the dataset into several consecutive buckets and assign each bucket a label. 
The problem is that it is hard to decide the length of the bucket for different query ranges. For example, when the query range is much larger than the bucket length, it is necessary to conduct separate ANN queries on many buckets, which causes reduced performance.

\begin{table}[!htbp]
    \centering
    \small
    \setlength{\tabcolsep}{0pt}
    \vspace{-2mm}
    \caption{Memory footprint (GB).}
    \vspace{-4mm}
    \begin{tabularx}{\columnwidth}{l *{5}{>{\centering\arraybackslash}X}}
    \toprule
         &  WIT & TripClick & Redcaps & YT-Rgb & YT-Audio\\
    \hline
    Raw Vectors & 7.7 & 2.9 & 2.0 & 3.9 & 0.47\\ 
    \hline
     \textbf{\texttt{iRangeGraph}}    &  12.87 & 4.34 & 7.14 & 9.05 & 1.95 \\
     \textbf{\texttt{SuperPostfiltering}} & 28.98 & 14.67 & 11.8 & 17.53 & 7.50\\
     \textbf{\texttt{Milvus}}& 8.38 & 3.23 & 2.21 & 4.28 & 0.66 \\
     \textbf{\texttt{Pre-filtering}} & 7.7 & 2.9 & 2.0 & 3.9 & 0.47\\ \hline
      \texttt{2DSegmentGraph} & 9.35 & 4.23 & 3.19 & 5.22 & 1.61\\
     \texttt{FilteredVamana} & 8.17 & 3.29 & 2.31 & 4.18 & 0.84\\
     \texttt{StitchedVamana} & 8.15 & 3.22 & 2.21 & 4.06 & 0.72\\
     \bottomrule
    \end{tabularx}
    \label{tab:index size}
\end{table}

\begin{table}[thbp]
    \centering
    \small
    \setlength{\tabcolsep}{0pt}
    \vspace{-3mm}
    \caption{Indexing time (s).}
    \vspace{-4mm}
    \begin{tabularx}{\columnwidth}{l *{5}{>{\centering\arraybackslash}X}}
    \toprule
         &  WIT & TripClick & Redcaps & YT-Rgb & YT-Audio\\
    \hline
     \textbf{\texttt{iRangeGraph}}    &  3,776  & 621 & 1,851  & 4,719 & 236 \\
     \textbf{\texttt{SuperPostfiltering}} & 11,603 & 5,140 & 1,765 & 4,735 & 1,206 \\
     \textbf{\texttt{Milvus}} & 784 & 476 & 174 & 593 & 49 \\
     \textbf{\texttt{Pre-filtering}}  & <10 & <10 & <10 & <10 & <10 \\ \hline 
     \texttt{2DSegmentGraph} & 16,865 & 3,165 & 1,412 & 3,038 & 381\\
     \texttt{FilteredVamana} & 358 & 144 & 80 & 110 & 35\\
     \texttt{StitchedVamana} & 450 & 54 & 41 & 85 & 16\\
     \bottomrule
    \end{tabularx}
    \vspace{-4mm}
    \label{tab:index time}
\end{table}

\subsubsection{The Results of the Ablation Study}
\label{subsec:exp ablation study}

\begin{figure*}
    \centering
    \includegraphics[width=\textwidth]{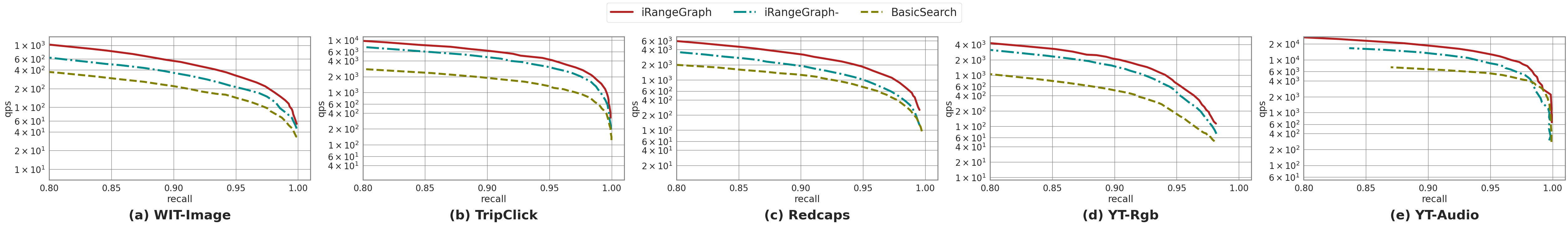}
    \vspace{-4mm}
    \caption{The ablation study of our core algorithm (constructing the dedicated graph on the fly) and edge selection algorithm~\ref{alg:edge selection}.}
    \vspace{-2mm}
    \label{fig:ablation}
\end{figure*}

In this section, we investigate the effect of the components in our search algorithm on the search performance. 
The qps-recall curves are presented in Figure~\ref{fig:ablation} with the query workloads of mixed range across all datasets. The ablation involves two folds. 

In order to evaluate the effectiveness of our core algorithm for constructing a dedicated graph for any query range for RFANN, we compare it with a trivial baseline, which we call \texttt{BasicSearch}. 
With the structure of segment tree, any query range can be expressed as the union of $O(\log n)$ non-overlapping segments in the tree. 
For example in Figure~\ref{fig:STMG}, the query range $[6,15]$ can be exactly expressed by the union of 5 disjoint segments, i.e., segment $[9,12]$ at L2, segments $[7,8],[13,14]$ at L3, and segments $[6],[15]$ at L4.
The \texttt{BasicSearch} baseline independently conducts ANN search on the elemental graphs of the $O(\log n)$ segments and finally merges the results.
We note that it is the way the segment tree is used for handling other types of range-based queries, e.g., range maximum query and range sum query~\cite{SegmentTree}. Figure~\ref{fig:ablation} shows that our method outperforms this trivial baseline by 2x to 4x in efficiency on all workloads at 0.9 recall.

We next evaluate the effectiveness of the efficient algorithm for edge selection. 
Note that the trivial algorithm of edge selection without skipping some layers has the time complexity of $O(m \log n)$ while the efficient algorithm has the amortized time complexity of $O(m + \log n)$ (Section~\ref{subsec:query}). 
The former introduces unignorable overhead in the search algorithm. 
According to Figure~\ref{fig:ablation}, we observe that the search performance based on the efficient algorithm (\texttt{iRangeGraph}, the red curve) 
achieves consistent improvement over the trivial algorithm (\texttt{iRangeGraph-}, the blue curve), which verifies the effectiveness of the efficient algorithm for edge selection.

\subsubsection{ Scalability Results.}
\label{subsec:exp scalability}

 We evaluate the scalability of our method with a larger public dataset DEEP~\footnote{\url{https://research.yandex.com/blog/benchmarks-for-billion-scale-similarity-search}}. The results show that as the dataset scales, the indexing cost remains moderate and the search performance remains promising. For more details, we refer to our technical report~\cite{technical_report}.

\subsubsection{ Comparison with \texttt{Oracle-HNSW}}
\label{subsec:exp oracle hnsw}

\begin{figure*}[thb]
    \centering
    \includegraphics[width=\textwidth]{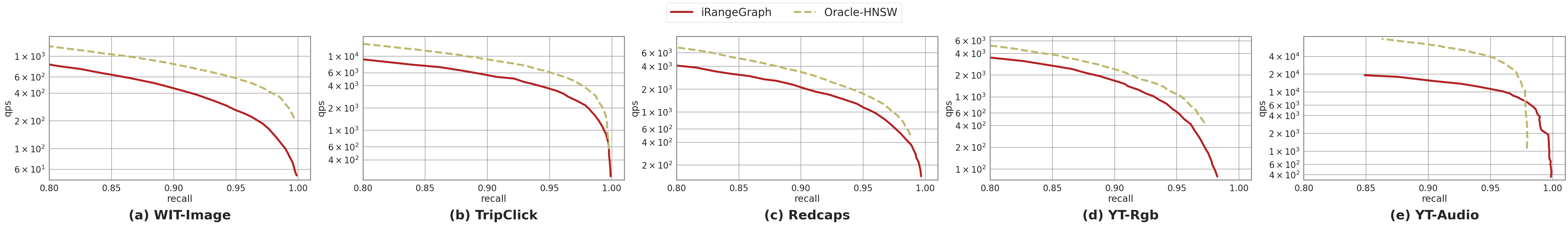}
    \vspace{-4mm}
    \caption{ Comparison between \texttt{iRangeGraph} and \texttt{Oracle-HNSW} under mixed range fraction.}
    \label{fig:oracle}
    \vspace{-2mm}
\end{figure*}

Here, we evaluate the performance gap between our dedicated graph and the graph that is explicitly materialized for the given query ranges. 
Specifically, in this experiment, we explicitly build an HNSW graph for every given query range and measure its search performance. We call this method \texttt{Oracle-HNSW}.
Note that \texttt{Oracle-HNSW} is not practical since it has space complexity $O(n^3m)$.
This experiment is only intended to quantify the performance gap between our method and the one that materializes all possible dedicated graphs for RFANN queries (in an impractical way).
In practice, query ranges are not known during the index phase, so it is necessary to materialize the HNSWs for all possible query ranges. Thus, the memory consumption of Oracle-HNSW in this experiment does not reflect its memory consumption in practice. We use a mixed query workload in this study. 
Specifically, we randomly generate a specific query range for each of the subset of 100 query vectors (but not a query range for a query as we did in Section~\ref{subsec:exp single attribute}); since otherwise, we would have to build 1,000 HNSWs for the 1,000 distinct query ranges, which renders the study infeasible.
According to Figure~\ref{fig:oracle}, on most of the datasets (except for YT-Audio), our practical algorithm has its search performance very close to \texttt{Oracle-HNSW}, e.g., \texttt{Oracle-HNSW} only outperforms our method by less than 2x in efficiency when reaching 0.9 recall.
On YT-Audio, \texttt{Oracle-HNSW} outperforms our method for many recall values, but not at very high recalls.
This might be explained by the fact that both our method and HNSW involve heuristic approximation of the RNG graph, and it is possible that a method does not produce stable performance across different datasets, as has been observed in a recent benchmark study~\cite{graph_benchmark}.

\subsubsection{The Results of the Multi-Attribute RFANN Query}
\label{subsec:exp multi attribute}

\begin{figure}
    \centering
    \includegraphics[width=\linewidth]{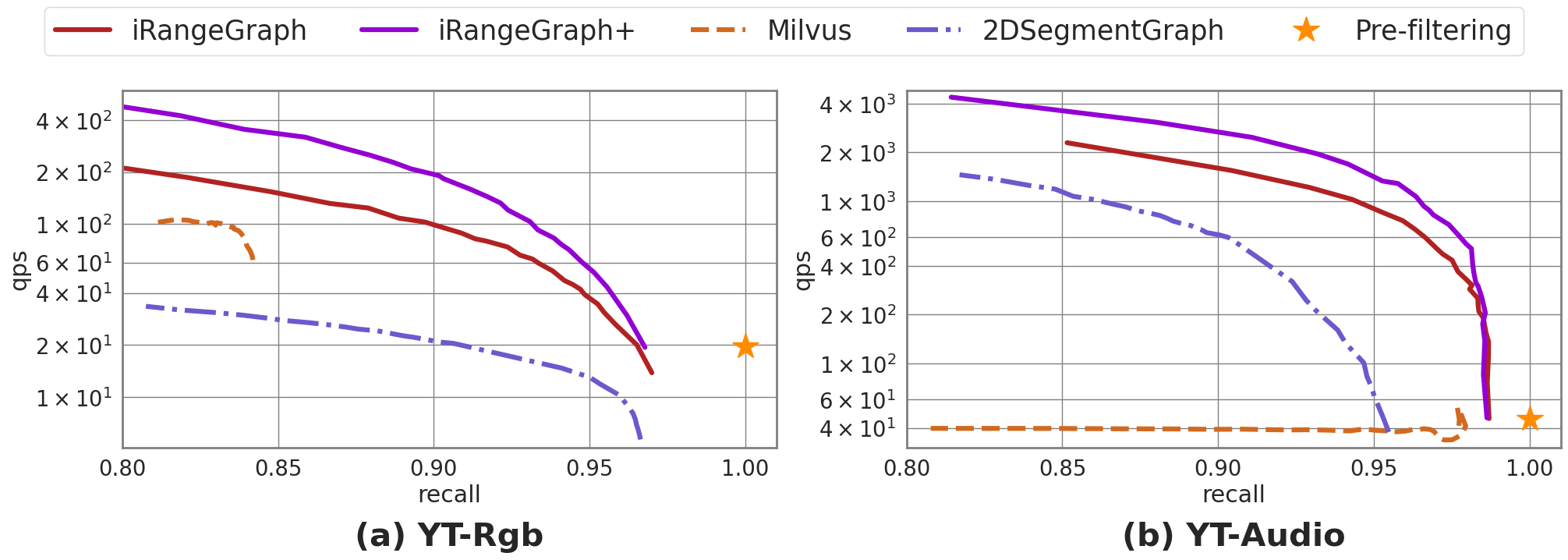}
    \vspace{-4mm}
    \caption{Multi-attribute RFANN query performance.}
    \vspace{-4mm}
    \label{fig:multi}
\end{figure}

This section assesses the effectiveness of the extension of \texttt{iRangeGraph} (Section~\ref{sec:MultiAttribute}) to the RFANN query with conjunctive predicates on multiple attributes.
Recall that the query aims to find the nearest neighbor that satisfies \textit{all} the constraints on the attributes. We note that it is likely that only a few data objects satisfy the conjunctive predicates (i.e., the selectivity of the conjunctive predicates is high), especially when the conjunctive predicates involve many attributes.
Note that this is an inherently simple workload for the query as it can be efficiently handled by the \texttt{Pre-filtering} strategy.
Therefore, we evaluate our method on non-trivial query workloads, where the selectivity of the conjunctive predicates is moderate. 
Specifically, note that each data object in the YT-Rgb and YT-Audio dataset involve two numeric attributes.
For each query vector, we randomly assign a query range with the expected range fraction of $2^{-2}$ to each of the attributes so that for the conjunctive predicates, the selectivity is moderate.
As for the baseline methods, we note that \texttt{SuperPostfiltering} and \texttt{Filtered-DiskANN} (including \texttt{FilteredVamana} and \texttt{StitchedVamana}) are not included in this experiment because they do not support the query with multiple range constraints.
Considering the baseline methods \texttt{2DSegmentGraph}, \texttt{Milvus}, and \texttt{Pre-filtering}, we note that they either provide extensions to the multi-attribute query (\texttt{2DSegmentGraph}~\cite{segmentgraph} and \texttt{Milvus}~\cite{wang2021milvus}) or can be extended to the query trivially (\texttt{Pre-filtering}).
For multi-attribute queries, \texttt{iRangeGraph} and \texttt{2DSegmentGraph} build indexes based on one of the attributes. For both methods, we build indexes based on the first attribute of the dataset (i.e., \# of likes for YT-Rgb and the publish time for YT-Audio)  and adopt \texttt{Post-filtering} for handling the second attribute (according to our experiments, using \texttt{In-filtering} in this case cannot achieve >0.8 recall). 
Besides, recall that we have proposed a simple idea (in Section \ref{sec:MultiAttribute}) that generalizes \texttt{In-filtering} and \texttt{Post-filtering} by allowing the search algorithm to visit the out-of-range neighbors with probability $p$. 
In order to prevent the algorithm from visiting too many out-of-range objects, $p$ is in practice set to $\exp(-t)$, where $t$ is the number of consecutive out-of-range objects that have been visited in the search path.
We evaluate the effectiveness of this technique by equipping \texttt{iRangeGraph} with it, denoted as \texttt{iRangeGraph+}.

As Figure \ref{fig:multi} indicates, (1) the extension of \texttt{iRangeGraph} can effectively cope with multi-attribute RFANN query workloads with moderate selectivity. It outperforms \texttt{2DSegmentGraph} by more than
2x speed-ups at 0.9 recall on both datasets. Further, \texttt{iRangeGraph} also outperforms \texttt{Milvus} with a much better qps-recall trade-off. (2) We note that the extension of \texttt{iRangeGraph} is not perfect for the multi-attribute RFANN query. It is shown that unlike \texttt{Pre-filtering}, it cannot achieve 1.0 recall on these datasets. However, \texttt{iRangeGraph} still provides a better qps-recall trade-off if the requirement on recall is relaxed, e.g., \texttt{iRangeGraph} offers 5x higher qps at 0.9 recall on YT-Rgb, and 35x higher qps at 0.9 recall on YT-Audio than \texttt{Pre-filtering}. (3) The idea of visiting out-of-range neighbors with probability $p$ helps enhance the search performance. For example, on both datasets, \texttt{iRangeGraph+} improves the qps by more than $70\%$ for \texttt{iRangeGraph} at 0.9 recall.

\section{related work}
\label{sec: related work}

\smallskip\noindent\textbf{Approximate Nearest Neighbor Search.} 
The existing methods for ANN query in high-dimensional Euclidean space can be generally divided into four categories: the graph-based, the quantization-based, the hashing-based and the tree-based. 
We refer readers to recent tutorials~\cite{tutorialThemis, tutorialXiao} and benchmarks/surveys~\cite{li2019approximate, annbenchmark, graph_benchmark, dobson2023scaling_billion_benchmark,wang2023graph,aumullerrecent2023} for a comprehensive review. 
Graph-based methods~\cite{fanng, nsw, fu2016efanna, iwasaki2016pruned, spann, rng, graph_benchmark, hnsw,nsg,2019diskann,li2019approximate,ChenW18,nssg} use a graph to connect data vectors and a heuristic search algorithm, usually greedy beam search, on the graph to find vectors that are close to query vector.
Among the graph-based methods, HNSW~\cite{hnsw}, NSG~\cite{nsg} and DiskANN~\cite{2019diskann}, which are all approximate RNG-based graphs~\cite{graph_benchmark}, have been widely used in industry~\cite{nsg,2019diskann,douze2024faiss,wang2021milvus}.
In our paper, we also build approximate RNG-based graphs as the indexes. Quantization-based methods~\cite{pq, pqfs, opq, addq, guo2020accelerating, babenko2014inverted,gao2024rabitq} accelerate searching by reducing the cost of distance computation and memory usage. 
Hashing-based methods~\cite{gan2012locality, datar2004locality, huang2015query, indyk1998approximate, sun2014srs, tao2010efficient} offer a theoretical guarantee on the probability of finding approximate nearest neighbors. Tree-based methods~\cite{beygelzimer2006cover, arya1993approximate, gu2022parallel} are powerful in searching the nearest neighbor in low-dimensional space, but they suffer from the curse of dimensionality.

\smallskip\noindent\textbf{Attribute-filtering ANN Search.}
Various algorithms and systems are developed for attribute-filtering ANN query~\cite{HQI,wang2021milvus,wei2020analyticdb,vearch,NHQ,HQANN,AIRSHIP,matsui2018reconfigurable,filtereddiskann,patel2024acorn,segmentgraph,WindowFilter,gupta2023caps}.
Besides the range-filtering ANN query~\cite{segmentgraph,WindowFilter}, there are also many studies targeting the ANN query with different attributes and predicates~\cite{filtereddiskann,AIRSHIP,gupta2023caps,NHQ,HQANN,HQI,matsui2018reconfigurable,patel2024acorn}. 
For example, Filtered-DiskANN~\cite{filtereddiskann} is developed for predicates on categorical attributes~\cite{filtereddiskann, NHQ, HQANN, AIRSHIP}. 
Some systems and algorithms (e.g., Milvus~\cite{wang2021milvus}) are developed to support predicates on generic attributes~\cite{HQI,patel2024acorn,wang2021milvus,vbase}.
Because different types of attributes involve highly diversified properties, for each type of attribute, it often entails specialized design in order to achieve the most competitive search performance on their corresponding attribute-filtering ANN queries.
Among those which support predicates on categorical attributes, we adapt the state-of-the-art method Filtered-DiskANN~\cite{filtereddiskann}  as the baseline in our experiments. Besides, the prevalent system Milvus~\cite{wang2021milvus} that supports predicates on generic attributes is also included in our experiments.
The experimental results (Section~\ref{subsec:exp single attribute}) show that they indeed incur suboptimal performance on the range-filtering ANN query compared with the specialized methods.
We note that HQI~\cite{HQI} is a method which further optimizes the generic attribute-filtering ANN query based on prior knowledge on the query workloads and conducts batched query execution offline, whose problem setting is different from ours. 
There are also studies which research on the attribute-filtering ANN query for multi-dimensional vectors~\cite{ferhatosmanoglu2001constrained}. Due to the curse of dimensionality, these methods cannot be easily adapted to the query for high-dimensional vectors and achieve competitive performance with the graph-based methods.

\smallskip\noindent\textbf{Similarity Search in General Metric Spaces.}
Beyond similarity search in Euclidean space, a large number of studies consider similarity search in general metric spaces~\cite{patella2008many,patella2009approximate,hjaltason2003index,chen2022indexing,zhu2022pivot,ciaccia2001approximate,kalantari1983data,dehne1987voronoi,uhlmann1991satisfying,navarro2002searching,chavez2014faster,chavez2016faster,ciaccia1998bulk,mtree,ruiz1986algorithm,movsko2011clustered,burkhard1973some,baeza1994proximity,yianilos1993data,chen2015efficient,florez2008hrg,foster2022generalized,chavez2006half,faloutsos1995fastmap,wang2005metricmap,cantone2005antipole,chavez2003probabilistic,zezula1998approximate,clarkson1997nearest}. In particular, a metric space is defined by a function with the properties of symmetry, positivity and triangle inequality. 
Euclidean space is an instance of metric spaces.
The techniques proposed for similarity search in metric spaces usually do not rely on the additional properties of a specific space.
Next, studies on similarity search in metric spaces can be generally divided into two threads, the exact and the approximate. 
The exact methods mostly conduct partitioning (of the metric space or dataset) and pruning (based on the triangle inequality)~\cite{kalantari1983data,dehne1987voronoi,uhlmann1991satisfying,navarro2002searching,chavez2014faster,chavez2016faster,ciaccia1998bulk,cantone2005antipole,mtree,ruiz1986algorithm,movsko2011clustered,burkhard1973some,baeza1994proximity,yianilos1993data,chen2015efficient,chavez2006half,florez2008hrg,foster2022generalized} to accelerate the search.
The approximate methods usually target better efficiency by allowing some sacrifice on accuracy~\cite{faloutsos1995fastmap,wang2005metricmap,cantone2005antipole,chavez2003probabilistic,zezula1998approximate,clarkson1997nearest}. 
In particular, we note that graph-based methods that are popular for the ANN query in Euclidean space are also being studied widely in general metric spaces~\cite{navarro2002searching,chavez2014faster,chavez2016faster,chavez2006half,florez2008hrg,foster2022generalized}.
In addition, in general metric spaces, attribute-filtering queries are also valuable as users may pose hard constraints on their targeted data objects. We note that the method proposed in this paper is oblivious to the specific data type and metric function.
This suggests that it has potential for being used in metric spaces as well (e.g., in combination with the graph-based methods).

\section{conclusion}
\label{sec: conclusion}

In conclusion, in this paper, for the range-filtering ANN query, we propose a method named \texttt{iRangeGraph}.
It constructs elemental graphs with moderate space consumption in the index phase and uses them to improvise range-dedicated graphs on the fly in the query phase.
The algorithm \texttt{iRangeGraph} is further extended to support multi-attribute range-filtering ANN query.
Extensive experiments on real-world datasets confirm its superior search performance over existing methods and moderate space consumption.

We would like to mention the following extensions and possible directions as future work.
(1) The memory footprint of \texttt{iRangeGraph} could be further reduced by building the index based on a multi-branch tree. As more branches lead to fewer layers of the index, and as each object appears once in a layer, the memory footprint will be further reduced when a multi-branch tree is used. 
(2) For the attribute-filtering ANN queries with other types of predicates and attributes, there are also well-established classical algorithms which can handle the filtering (e.g., segment tree can handle the range filtering).
It is worth exploring to extend the idea of \texttt{iRangeGraph} to the attribute-filtering for other types of attributes and predicates. 
(3) It would be an interesting research topic to further explore the dynamic updates, insertion and deletion of the \texttt{iRangeGraph} index given its promising search performance.

\section*{Acknowledgements}
We would like to thank the anonymous reviewers for providing constructive feedback and valuable suggestions. This research is supported by the Ministry of Education, Singapore, under its Academic Research Fund (Tier 2 Award MOE-T2EP20221-0013, Tier 2 Award MOE-T2EP20220-0011, and Tier 1 Award (RG77/21)). This research is also supported by the Innovation Fund Denmark centre, DIREC. C- S. Jensen was supported in part by the Innovation Fund Denmark centre, DIREC. Any opinions, findings and conclusions or recommendations expressed in this material are those of the author(s) and do not reflect the views of the Ministry of Education, Singapore.

\bibliographystyle{ACM-Reference-Format}
\bibliography{main}

\end{document}